\documentclass{aa}

\usepackage{graphicx}
\usepackage{txfonts}
\usepackage[colorlinks=true,linkcolor=blue,citecolor=blue,filecolor=blue,urlcolor=blue]{hyperref} 
\usepackage{natbib}
\usepackage[title]{appendix}
\usepackage{subfig}

\usepackage{xspace}
\usepackage{xcolor}

\usepackage[export]{adjustbox}

\defcitealias{Belloni2017a}{Belloni et~al.}
\defcitealias{Belloni2017b}{Belloni et~al.}
\newcommand\citetBab[2]{\citeauthor{#1}\ (\citeyear{#1}a, b)}

\defcitealias{Cordoni2020a}{Cordoni et~al.}
\defcitealias{Cordoni2020b}{Cordoni et~al.}
\newcommand\citetCab[2]{\citeauthor{#1}\ (\citeyear{#1}a, b)}

\usepackage{xspace}
\usepackage{xcolor}

\usepackage[export]{adjustbox}

\newcommand{\CONCPOP}{$\text{conc}_{\text{pop}}$\xspace}

\newcommand{\MOCCA}{\textsc{mocca}\xspace}
\newcommand{\MOCCASURVEY}{\textsc{mocca-survey}\xspace}
\newcommand{\MOCCASURVEYTWO}{\textsc{mocca-survey-2}\xspace}
\newcommand{\BEANS}{\textsc{beans}\xspace}
\newcommand{\NBODY}{\textsc{nbody}\xspace}

\newcommand{\SURVEY}{\textsc{mocca-survey}\xspace}

\newcommand{\rhob}{$\rm r_{hl}$\xspace}

\newcommand{\RC}{$\rm r_{c}$\xspace}
\newcommand{\NSGNTOT}{$\rm N_{SG}/N_{TOT}$\xspace}
\newcommand{\msun}{$\rm M_{\odot}$\xspace}
\newcommand{\Mmax}{$\rm M_{max}$\xspace}
\newcommand{\rhfg}{$\rm r_{hFG}$\xspace}
\newcommand{\tdiss}{$\rm t_{diss}$\xspace}
\newcommand{\up}{$\uparrow$\xspace}
\newcommand{\down}{$\downarrow$\xspace}
\newcommand{\same}{$\approx$\xspace}
\newcommand{\upweak}{$\rm \uparrow_{w}$\xspace}
\newcommand{\downweak}{$\rm \downarrow_{w}$\xspace}

\newcommand{\upstrong}{$\rm \uparrow_{s}$\xspace}
\newcommand{\downstrong}{$\rm \downarrow_{s}$\xspace}

\newcommand{\rg}{$\rm r_{g}$\xspace}
\newcommand{\Wo}{$\rm W_0$\xspace}
\newcommand{\Wofg}{$\rm W_{0,FG}$\xspace}
\newcommand{\Wosg}{$\rm W_{0,SG}$\xspace}
\newcommand{\rtid}{$\rm r_{tid}$\xspace}

\newcommand{\kms}{{\rm {km\, s^{-1}}}}

\titlerunning{Evolution of multiple populations}

\begin{document}

   \title{MOCCA: Global properties of tidally filling and underfilling globular star clusters with multiple stellar populations}

   \subtitle{}

   \author{A. Hypki\inst{1,2}
          \and
          E. Vesperini\inst{3}
          \and
          M. Giersz\inst{2}
          \and
          J. Hong\inst{4}
          \and
          A. Askar\inst{2}
        \and
        M. Otulakowska-Hypka\inst{5}
        \and
        L. Hellstrom\inst{2}
        \and
        G. Wiktorowicz\inst{2}
          }

   \institute{Faculty of Mathematics and Computer Science, A. Mickiewicz
University, Uniwersytetu Pozna\'nskiego 4, 61-614 Pozna\'n, Poland
         \and
             Nicolaus Copernicus Astronomical Center, Polish Academy of Sciences, Bartycka 18, 00-716 Warsaw, Poland\\
          \email{ahypki@camk.edu.pl}
         \and
            Indiana University Department of Astronomy, 727 East Third Street, Bloomington, IN 47405, USA
        \and
            Korea Astronomy and Space Science Institute, Daejeon 34055, Republic of Korea
        \and
            Astronomical Observatory Institute, Faculty of Physics, A. Mickiewicz
University, S\l{}oneczna 36, 60-286 Pozna\'n, Poland
             }

   \date{Received xx; accepted xx}

 
  \abstract
  {We explore the evolution of various properties of multiple-population globular clusters (GCs) for a broad range of initial conditions. We simulated over 200 GC models using the \MOCCA Monte Carlo code  and find that  present-day properties (core and half-light radii, ratio of the number of second-generation (SG) stars to the total number of stars, \NSGNTOT) of these models cover the observed values of these quantities for Milky Way GCs. Starting with a  relatively small value of the SG fraction (\NSGNTOT $\sim 0.25$) and a SG system concentrated in the inner regions of the cluster, we find, in agreement with previous studies, that systems in which the first-generation (FG) is initially tidally filling or slightly tidally underfilling  best reproduce the observed ratios of \NSGNTOT and have values of the core and half-light radii typical of those of many Galactic globular clusters. Models in which the FG is initially tidally underfilling retain values of \NSGNTOT close to their initial values. These simulations expand previous investigations and serve to further constrain the viable range of initial parameters and better understand their influence on present-day GC properties. The results of this investigation also provide the basis for our future survey aimed at building specific models to reproduce the observed trends (or lack thereof) between the properties of multiple stellar populations and other clusters properties.}


   \maketitle
%

\section{Introduction}
\label{s:Intro}

Globular clusters were previously believed to be simple stellar populations with a uniform age and chemical composition. However, thanks to extensive photometric and spectroscopic studies, it has become clear that these systems are more complex than previously thought and host multiple stellar populations (MSP) characterized by differences in the chemical abundances of various light elements and, in some cases, also of iron (see e.g. \citealt{Gratton2019A&ARv..27....8G} for a review). This discovery has opened many new questions and challenged the traditional understanding of globular cluster formation and dynamical evolution.

The origin of MSP is still matter of intense investigation and no consensus has been reached on the possible sources of processed gas necessary to explain the  chemical composition of MSP and their star formation history. Some MSP formation models have proposed that all stars formed simultaneously, but some stars acquired an anomalous chemical composition by accreting processed gas produced by supermassive stars or massive binary stars (see e.g.,\citealt{2009A&A...507L...1D},\citealt{Bastian2013}, \citealt{Gieles2018}). 

Other formation models suggest that after the formation of first generation stars (hereafter FG), a second episode of star formation takes place from gas enriched by the ejecta of FG polluters and this results in the production of second-generation stars (hereafter SG) with different chemical abundances (e.g. enhancement in Na, Al, N abundances and depletion in Mg, O, and C). In this scenario, the origin of the gas that formed later generations of stars is of particular interest: possible candidates for the sources of polluted gas proposed in the literature include single and binary Asymptotic Giant Branch (AGB) stars, rapidly rotating massive stars, massive interacting binaries, stellar mergers, black hole accretion disks (see e.g. \citealt{Ventura2001}, \citealt{Decressin2007}, \citealt{deMink2009}, \citealt{Elmegreen2017}, \citealt{Krause2013}, \citealt{DErcole2008MNRAS.391..825D,DErcole2010}, \citealt{Jerabkova2017}, \citealt{DAntona2016}, \citealt{Breen2018}, \citealt{Calura2019MNRAS.489.3269C}, \citealt{Wang2020MNRAS.491..440W}).

While the origin of MSP in globular clusters is not fully understood, different proposed scenarios based either on numerical hydrodynamical simulations (see e.g. \citealt{DErcole2008MNRAS.391..825D}, \citealt{Bekki2010ApJ...724L..99B,Bekki2011MNRAS.412.2241B}, \citealt{Calura2019MNRAS.489.3269C}, \citealt{Lacchin2022}) or general considerations (see e.g. \citealt{Gieles2018}) typically share the prediction that SG stars tend to form more centrally concentrated compared to FG stars.

Although this difference in the initial spatial distributions of FG and SG stars is gradually erased during the cluster dynamical evolution, some clusters may still retain some memory of the structural properties imprinted by the formation process (see e.g. \citealt{Vesperini2013}, \citealt{Miholics2015}, \citealt{Vesperini2021MNRAS.502.4290V}, \citealt{Sollima2021MNRAS.502.1974S} for some studies of the process of spatial mixing) and, indeed, several observational studies have found evidence of SG stars being more centrally concentrated than the FG population (see e.g. \citealt{Bellini2009}, \citealt{Lardo2011}, \citealt{Simioni2016}, \citealt{Dalessandro2019ApJ...884L..24D}, \citealt{Onorato2023},\citealt{Lacchin2023}). See also \citet{Leitinger2023} for a study finding two clusters (NGC~3201, NGC~6101) where the FG is currently more centrally concentrated than the SG, but see \citet{Cadelano2024} for an observational study of the structural and kinematic properties of one of those clusters, NGC~3201, providing  support to scenarios in which the SG formed more centrally concentrated.

Differences between FG and SG stars are not limited to their spatial distributions but extend to their kinematic properties.
Differences in the FG and SG kinematic properties may be imprinted during the formation process (see e.g. \citealt{Bekki2010ApJ...724L..99B,Bekki2011MNRAS.412.2241B}, \citealt{Lacchin2022} for differences in the FG and SG rotation) or emerge during the cluster evolution (see e.g. \citealt{Vesperini2021MNRAS.502.4290V} for differences between the anisotropy in the FG and SG velocity distribution).
Evidence of kinematic differences have been found in several observational studies (see e.g. \citealt{Richer2013ApJ...771L..15R}, \citealt{Bellini2015}, \citealt{Cordero2017MNRAS.465.3515C}, \citetCab{Cordoni2020a}{Cordoni2020b}, \citealt{Libralato2023}).

The initial structural differences between the FG and SG populations may also have important implications for the survival and evolution of their binary stars.
\citet{Hong2015MNRAS.449..629H,Hong2016MNRAS.457.4507H,Hong2019MNRAS.483.2592H} investigated the evolution of binary stars in MSP through direct \NBODY simulations. As a consequence of the fact that the SG population is initially more centrally concentrated in a denser subsystem, SG binaries are either more easily disrupted compared to FG binaries or, for compact binaries, they evolve more rapidly towards more compact configurations. These investigations also predicted the presence of mixed binaries, composed of stars from different generations due to member exchange during strong interactions.
Further detailed investigations of the dynamics of binary stars in multiple-population clusters and what they can reveal about the cluster initial structural properties have been carried out by \citet{Hypki2022} and \citet{Sollima2022}.

On the observational side, the investigation of the binary populations in MSP is still in its early stages. The findings of the few studies carried out so far are in general agreement with the predictions of numerical simulations concerning the fraction of FG and SG binaries and their variation with the cluster centric distance (see \citealt{DOrazi2010}, \citealt{Lucatello2015}, \citealt{Dalessandro2018ApJ...864...33D}, \citealt{Milone2020MNRAS.491..515M}). Possible evidence of primordial differences in the fraction of FG and SG binaries have been suggested in the study by \citep{Kamann2020}. The first evidence of the mixed binaries predicted by \citet{Hong2015MNRAS.449..629H} has been reported by \citet{Milone2020MNRAS.491..515M}.

In this paper we continue our previous investigations of the dynamics of MSP with an extensive survey of Monte Carlo simulations exploring a broad range of different initial conditions and shedding further light on the complex dynamics of MSP clusters.  The goal of this paper is to carry out a general exploration of how the evolution of some of the fundamental MSP properties such as the fraction of SG stars and of some of the clusters' structural properties depend on the cluster initial conditions. We emphasize that although in some cases we will compare our results to observations, our analysis is not specifically aimed at reproducing the observed trends and distributions of the clusters' observed properties; our general goal is rather to explore what initial conditions eventually lead to present-day properties within the range of those found in Galactic clusters, what their dynamical history is, and expand the theoretical framework describing how various parameters affect the dynamics of MSP. The conclusions of this work will help to design the initial conditions for the next, more comprehensive \MOCCASURVEY. A more extended survey and specific choices on the distribution of initial properties are required for a more detailed comparison with observations and will be the subject of future papers.

This paper is organized as following. In the Section~\ref{s:NumSimul} there is described the newest version of the \MOCCA code, initial conditions of the numerical simulations performed for this paper, and finally a short description about the data analysis software. Section~\ref{s:Results} presents the results obtained with the \MOCCA simulations through the context of the ratio between then number of objects from the SG to the total (\NSGNTOT). The \MOCCA simulations are also briefly compared with the Milky Way (MW) GCs showing that \MOCCA simulations are able to cover the observational ranges of cluster global properties and thus are good probes of the physical processes taking place between MSP too. In Section~\ref{s:Discussion} we discuss the potential implications obtained from \MOCCA simulations for the observational signatures of the multiple populations in GCs and some implications for the scenarios of their formation. Section~\ref{s:Conclusions} briefly summarizes main paper findings.

\section{Numerical simulations}
\label{s:NumSimul}

This section presents the description of the \MOCCA code, the simulations which were computed for this project and the way the output data were analyzed.

\subsection{MOCCA}
\label{s:MOCCA}

This work is based on the numerical simulations performed with the \MOCCA\footnote{\url{https://moccacode.net}} Monte Carlo code \citep{Giersz1998MNRAS.298.1239G,Hypki2013MNRAS.429.1221H,Giersz2014arXiv1411.7603G,Hypki2022}. \MOCCA is a feature-rich, advanced code that performs full stellar and dynamical evolution of real size star clusters. Over the last few years it has been substantially updated and many new features were added to the code which made it one of the most advanced and fastest codes in stellar dynamics able to simulate real-size star clusters all up to Hubble time. The newest major additions include several features to support the study of the dynamics and stellar evolution of multiple stellar populations. A detailed description of the new \MOCCA features was presented in \citet{Hypki2022}.

\MOCCA is able to follow the full dynamical evolution of MSP and also the stellar evolution for different populations. Only mergers and mass transfers between stars from different populations are treated in a simplified way -- the stars are marked as a mixed population, because we do not provide procedures to accurately model the chemical mixing between two stars belonging to different populations.

The initial conditions explored in this paper and in our previous studies (\citealt{Vesperini2021MNRAS.502.4290V}, \citealt{Hypki2022}) are inferred from the results of hydrodynamical simulations (see e.g. \citealt{DErcole2008MNRAS.391..825D}, \citealt{Calura2019MNRAS.489.3269C}) of multiple population formation from the ejecta of AGB stars and external pristine gas reaccreted by the cluster showing that SG stars form in a centrally concentrated subsystem embedded in a more extended FG system. As pointed out in Section~\ref{s:Intro}, however, the prediction of such a spatial configuration is generally shared also by other models based on different FG polluters.

Our simulations starts with the FG and SG subsystems already in virial equilibriums and do not follow in detail the very early phases of SG formation. \MOCCA allows for stellar evolution to start for all populations at T~=~0 or the stellar evolution of the SG can start after some time delay. New physics which was added to the stellar evolution part of the code (additions to \textsc{sse/bse} code \citep{Hurley2000MNRAS.315..543H, Hurley2002MNRAS.329..897H}, \citetBab{Belloni2017a}) is included. A comprehensive summary of the stellar evolution features of the \MOCCA code also can be found in \citet{Kamlah2021}. Strong dynamical interactions in \MOCCA are performed with \textsc{fewbody} code \citep{Fregeau2004-01-004,Fregeau2007ApJ...658.1047F}. The dissipative effects connected with tidal forces or gravitational wave radiation during dynamical \textsc{fewbody} interactions are not taken into account yet (they are planned for the next version of the code).

\subsection{Initial conditions}
\label{s:InitCond}

In this paper we extend the initial conditions explored in our previous studies and carry out a comprehensive investigation of the dependence of the evolution of the fraction of SG stars and the cluster structural parameters on the initial conditions.

Following previous works \citep[see][and references therein]{Hong2016MNRAS.457.4507H,Vesperini2021MNRAS.502.4290V,Hypki2022}, the initial model contains a higher number of FG stars and the initial number ratio between the SG/FG stars is typically set between 0.33 to 0.38. It is also assumed that the SG is more centrally concentrated than the FG population, and we use  the concentration parameter (\CONCPOP), defined as the ratio between the half-mass radii of SG to FG, to quantify the initial spatial differences between the two populations. Furthermore, in the initial model, the \citet{King1966AJ.....71...64K} concentration parameter ($\rm W_{0}$) is specified separately for each population. For the SG, this was fixed to \Wosg~=~7 or 8, and for the FG population this parameter was varied. For FG stars, initial zero-age main sequence (ZAMS) masses were sampled between 0.08 to 150 \msun. For SG stars, the upper limit for ZAMS mass was set to 20 \msun or 150 \msun. The ZAMS masses were sampled according to the \citet{Kroupa2001MNRAS.322..231K} initial mass function (IMF).

We explore the evolution of models at a various fixed Galactocentric distances, \rg, as indicated in Table~\ref{t:InitCond}.

Similar to the models simulated in \citet{Hypki2022}, the metallicity of both populations in all the simulated clusters models was set to $Z=0.001$ (5 per cent of $Z_{\rm \odot}$). All models were simulated with an updated treatment for the evolution of massive stars \citep{tanikawa2020,Kamlah2021} with improved treatment for mass loss due to stellar winds and the inclusion of pair and pulsational pair-instability supernova \citep{Belczynski2016A&A...594A..97B}. The masses of black holes (BH) and neutron stars (NS) were determined according to the rapid supernovae prescriptions from \citet{Fryer2012}. NS natal kicks were sampled from a Maxwelllian distribution with $\sigma = 265 \ \kms$ \citep{hobbs2005}. However, for BHs, these natal kicks were reduced according to the mass fallback prescription \citep{Belczynski_2002,Fryer2012}. The formation of neutron stars with negligible natal kicks through electron-capture supernova was also enabled \citep{Kamlah2021}. Another feature of these models is the inclusion of gravitational wave (GW) recoil kicks whenever two BHs merge \citep{baker2008,morawski2018}. The magnitude of the GW recoil kick depends on the magnitude and orientation of the spins of BHs. In all the simulated models, low birth spins for BHs are assumed and are uniformly sampled values between 0 and 0.1 \citep{fuller2019}. The orientation of the BH spin with respect to the binary orbit is randomly distributed \citep{morawski2018}. 

The initial conditions are summarized in the Table.~\ref{t:InitCond}. From now on the models are referred as \SURVEY. The table summarizes the initial conditions with all of the possible parameters. However, it is important to note, that we did not compute all possible combinations of these parameters but only a small subset of them. We computed over 200 \MOCCA models for the purpose of this paper.  

\begin{table}
\caption[Initial conditions for \SURVEY]{The table shows selected 
initial conditions of \MOCCA models, shortly \SURVEY, performed with the upgraded version of the code. All of the initial conditions were carefully chosen to test how to prolongate the GCs lifetimes, but with the goal of being able to reproduce the parameters of MW GCs (e.g. \rhob, core radii (\RC) total masses, \NSGNTOT) and to help to design the initial conditions for the next, more comprehensive \MOCCASURVEY. Some of the parameters are different for two stellar populations (FG, SG), e.g. N, and some of them are the same in both cases, e.g. \rg. The meaning of the initial parameters is following: 
     $\rm N$ -- initial number of objects; 
     \Wo -- King model parameter; 
     $\rm M_{max}$ -- upper mass limit for a single star [$\rm M_{\odot}$];
     $\rm fb$ -- binary fraction;
     \rg -- Galactocentric distance [kpc]; 
     \rhfg -- half-mass radius for the first population [pc]; 
     TF -- tidally filling model;
     \CONCPOP -- concentration parameter between two populations (e.g. value 0.1 means that SG has half-mass radius ($\rm R_h$) 10 times smaller than FG $\rm R_h$). We did not compute all possible combinations of these parameters, only a subset of them.}
\centering
\begin{tabular}{|c| c |c|}
\hline 
Parameter & FG & SG\\ 
\hline\hline
 $\rm N$$^{1}$         & 400k, 600k, 800k, 1.6M$^{2}$  & 150k, 200k, 300k, 600k \\ 
 $\rm W_{0}$     & 2, 3, 4, 5, 6             & 7, 8$^{3}$             \\
 \Mmax [$\rm M_{\odot}$]      & 150 & 20, 150$^{4}$       \\
 \textit{fb}     & \multicolumn{2}{|c|}{0.1, 0.5, 0.95} \\
 \rg [kpc] & \multicolumn{2}{|c|}{2, 4, 6, 8} \\
 \rhfg [pc] & \multicolumn{2}{|c|}{2, 4, 6, 8, TF} \\
 \CONCPOP & \multicolumn{2}{|c|}{0.05, 0.1, 0.2} \\
 \hline
    \multicolumn{3}{l}{$^{1}$\footnotesize{400k+150k, 600k+200k, 800k+300k, 1.6M+600k models only}} \\
    \multicolumn{3}{l}{$^{2}$\footnotesize{only for \rg~=~2, \Wofg~=~2,3}} \\
    \multicolumn{3}{l}{$^{3}$\footnotesize{only for \Wofg~=~2, 3}} \\
    \multicolumn{3}{l}{$^{4}$\footnotesize{only for N~=~600k+300k}} \\
\end{tabular}
\label{t:InitCond}
\end{table}

\subsection{Data analysis}
\label{s:DataAnalysis}

Data analysis on this paper was done in \BEANS\footnote{\url{https://beanscode.net}} software \citep{Hypki2018}. More precisely, the data analysis was performed in Apache Pig\footnote{\url{https://pig.apache.org}}, which is a high level language for Apache Hadoop\footnote{\url{https://hadoop.apache.org}} platform. \BEANS allows to have one script which is able to query various \MOCCA simulations (from different surveys) in bulk and analyze them in a distributed way. Next, they can be shared and redistributed among the collaborators for further analysis. One example Apache Pig script was discussed in \citet[Appendix B]{Hypki2022}, and another is discussed in the Appendix~\ref{s:AppendixB} of this paper.

\section{Results}
\label{s:Results}

In this section we describe the results obtained from the \MOCCA simulations while investigating some of structural properties of our models, and the evolution of \NSGNTOT.

We emphasize that our \MOCCA models are not aimed at reproducing the detailed properties of any specific globular cluster or the trends and correlations observed in the Galactic globular cluster system. In this paper, our focus is on the study of the evolution of multiple-population clusters and how the fraction of SG stars and the clusters' structural parameters depend on various initial properties. This will help to better constrain the initial conditions for a new \MOCCASURVEYTWO containing a few thousands of GC models. Although, in some figures we will plot the final values of some of the fundamental properties of our models along with the corresponding observed values of Galactic GCs, the goal will be just to provide a general comparison of the range of final properties of our models with the corresponding observed values.

\subsection{Structural properties}
\begin{figure}
\begin{center}
  \includegraphics[width=0.99\linewidth]{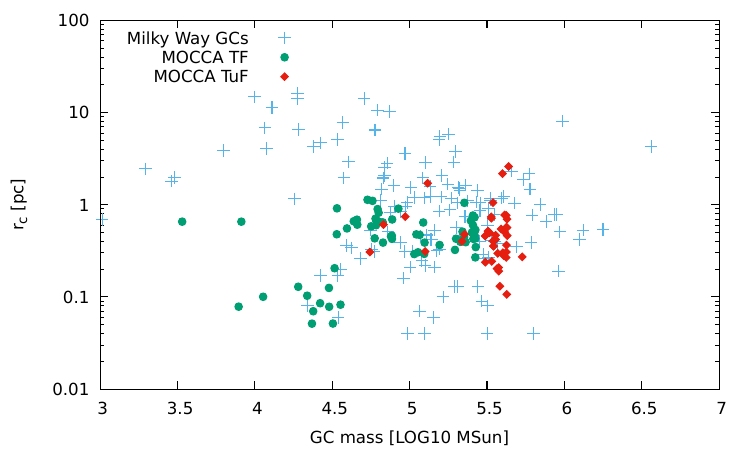}
  \includegraphics[width=0.99\linewidth]{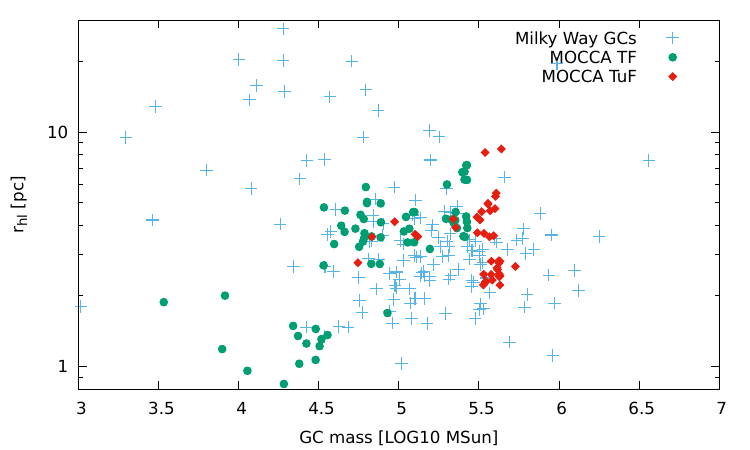}
  \includegraphics[width=0.99\linewidth]{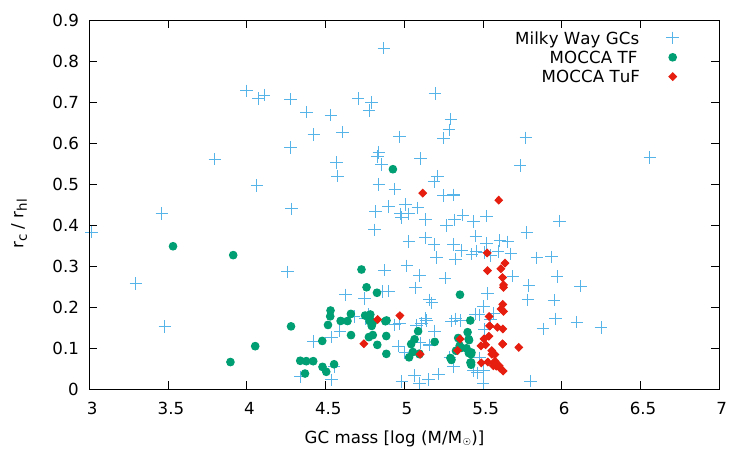}
  \caption{Milky Way GCs coverage with \MOCCA models for core (\RC), half-light (\rhob) radii, and the ratio between them (from the top to the bottom, respectively). Half-light and core radii for MW GCs are taken from \citet{Baumgardt2019MNRAS.482.5138B}.  Only \MOCCA models with the age at least 10 Gyr are shown.
  }
  \label{f:1:MWcoverage}
\end{center}
\end{figure}

In the three panels of Figure~\ref{f:1:MWcoverage}, we show the final values of core radius (\RC, a distance at which surface brightness is equal to half of the central one), half-light radii (\rhob) and the ratio \RC/\rhob for all the \MOCCA models which survived at least 10 Gyr as function of the final cluster's mass. 

The three panels also show the observed values for Galactic GCs that were taken from \citet{Baumgardt2019MNRAS.482.5138B}. While our models were not specifically designed to replicate the properties of the Galactic GC system, it is interesting to note how both, tidally filling (TF) and tidally underfilling (TuF) \MOCCA models, cover the range of values of the radii -- in general they do fall within the range of observed values. As already pointed out, exploration of a broader range of initial conditions is necessary to identify which models can evolve and reach values of the structural parameters not covered by our current survey. For example, larger values of  \RC and \rhob may be attained by systems evolving at larger Galactocentric distances than those studied in our survey and larger final values of the clusters' masses may be obtained for systems initially more massive than those we have considered.

\subsection{Evolution of \NSGNTOT ratios}
\label{s:EvNSGNTOT}
In this section we explore the evolution of the \NSGNTOT ratio and the clusters' half-light radii and how these depend on various initial properties of the cluster models. In Figure~\ref{f:2:NSG1} we show the time evolution of these two quantities for various representative cases with initial conditions indicated in each panel of this figure. Overall the panels of Figure~\ref{f:2:NSG1} provide a view of the role of various parameters on the evolution of \NSGNTOT and show that in all cases the \rhob values are generally consistent with those found in Galactic clusters.

We divide the cases presented in this figure into two groups: TF and TuF clusters and we discuss the role played by each of the parameters varied in our exploration.

The two panels in the top row of Figure~\ref{f:2:NSG1} show the evolution of TF and TuF models with different total number of stars. The TF models all evolve very similarly reaching large values of \NSGNTOT of $\sim 0.8$ and half-light radii of $\sim 4$ pc consistent with those observed in many Galactic clusters.
The most distinct difference are the longer dissolution times for models with larger N: this trend is simply the consequence of the larger masses and accordingly larger half-mass relaxation times and slower cluster evolution.

The TuF models also evolve towards final half-light radii of 3-4 pc while the final \NSGNTOT are much smaller than those found in the TF models and are just slightly smaller than the initial ones ($\sim 0.2$). It is expected because these clusters undergo a weaker early loss of FG stars resulting in smaller final values of \NSGNTOT, although they still fall in the range of values observed in Galactic clusters.

\begin{figure*}
\begin{center}
  \includegraphics[width=0.49\linewidth]{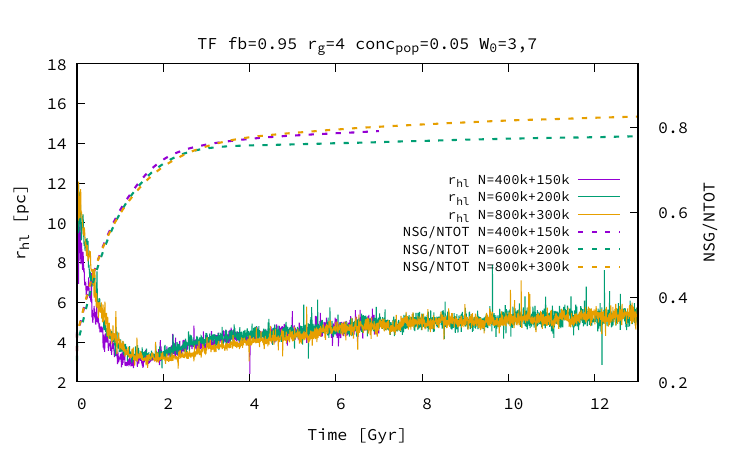}
  \includegraphics[width=0.49\linewidth]{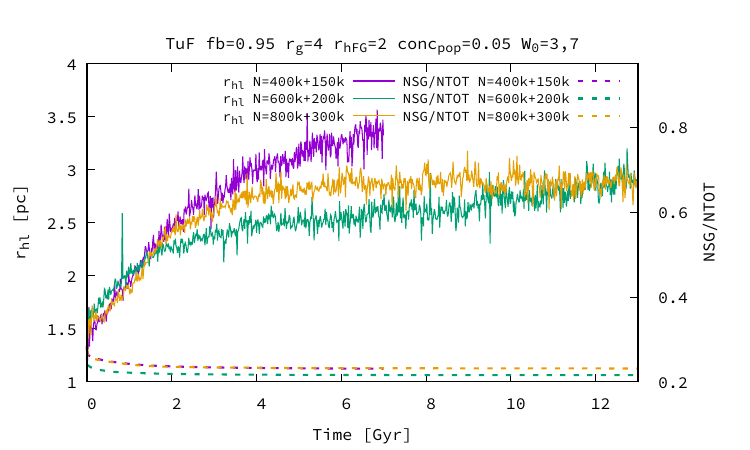}
  \includegraphics[width=0.49\linewidth]{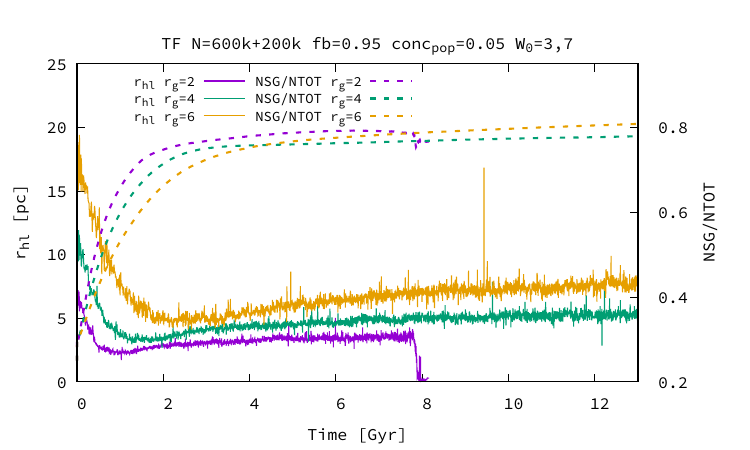}
  \includegraphics[width=0.49\linewidth]{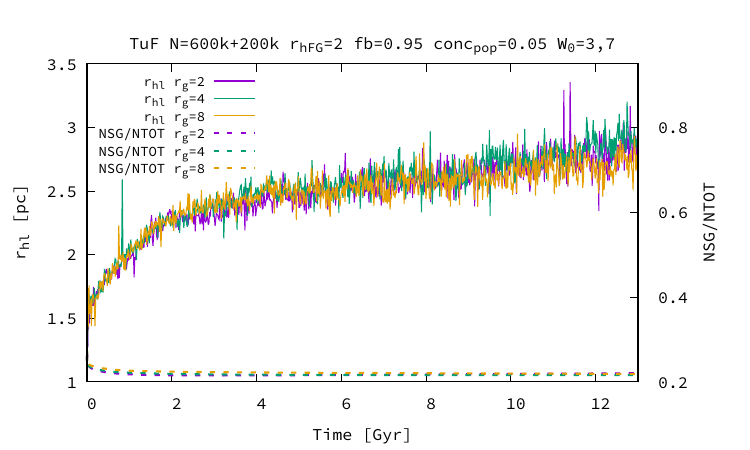}
  \includegraphics[width=0.49\linewidth]{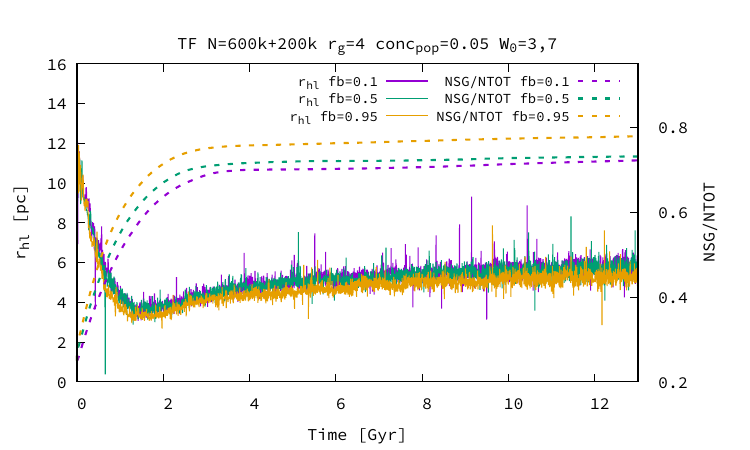}
  \includegraphics[width=0.49\linewidth]{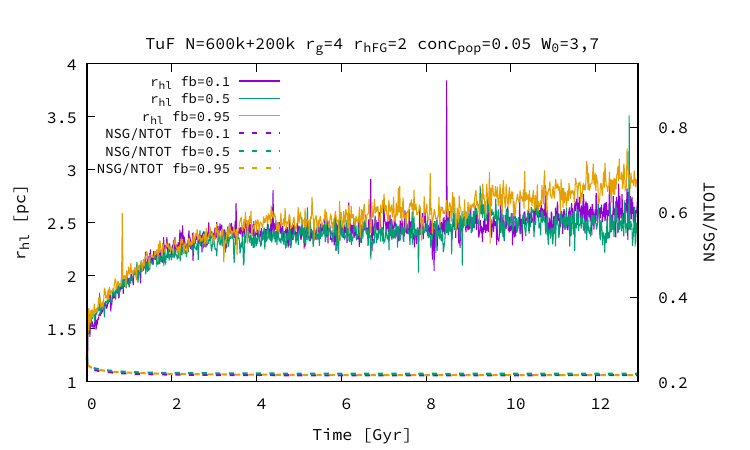}
  \includegraphics[width=0.49\linewidth]{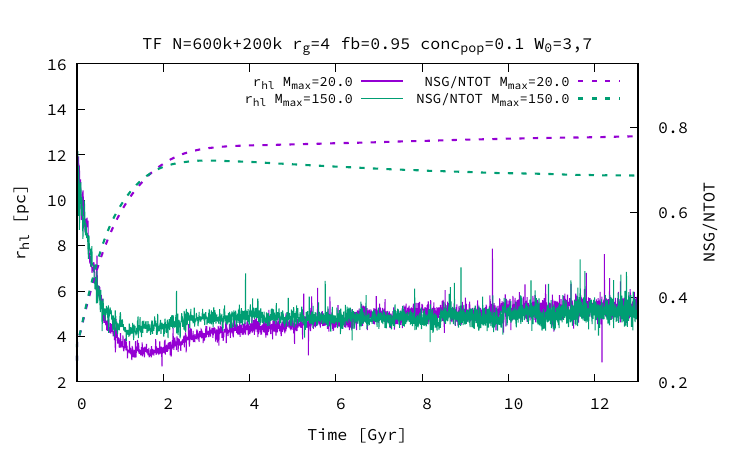}
  \includegraphics[width=0.49\linewidth]{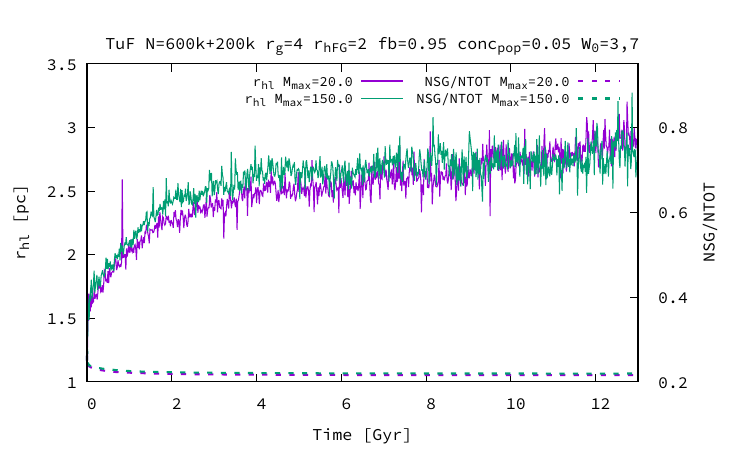}
  \caption{Evolution of \NSGNTOT ratios, together with half-light radii (\rhob), for a set of \MOCCA simulations. Every row consists of two series of \MOCCA models: tidally filling (left, TF), and tidally underfilling (right, TuF). The main \MOCCA initial parameters are summarized in the titles. On every plot there are 2-4 \MOCCA models for which all initial parameters are the same, except one shown in the legend of each panel. In each panel,  the left Y axis shows the half-light radius (\rhob [pc]), and the right Y axis the \NSGNTOT ratio. Starting from the top, each row shows models differing in:  $N$ -- initial number of stars , \rg [kpc] -- Galactocentric distance, fb -- binary fraction, and \Mmax [\msun] -- maximum mass for the SG stars.
  }
  \label{f:2:NSG1}
\end{center}
\end{figure*}

\begin{figure*}
  \ContinuedFloat 
\begin{center}
  \includegraphics[width=0.49\linewidth]{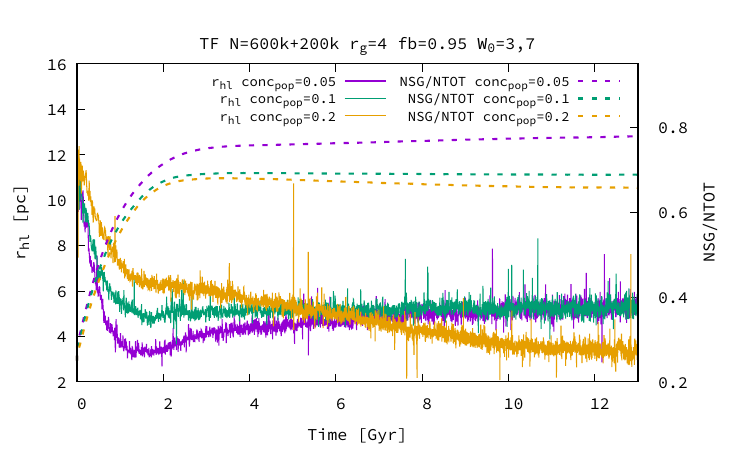}
  \includegraphics[width=0.49\linewidth]{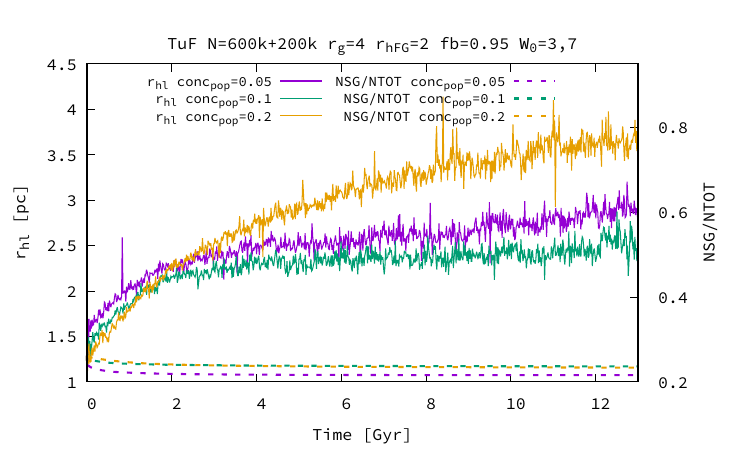}
  \includegraphics[width=0.49\linewidth]{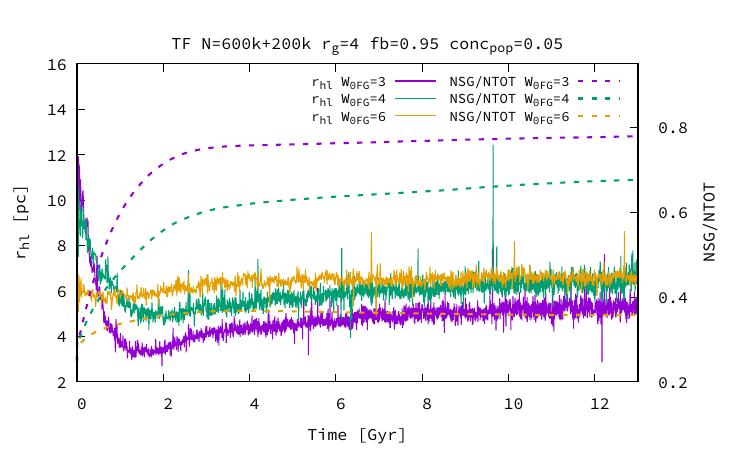}
  \includegraphics[width=0.49\linewidth]{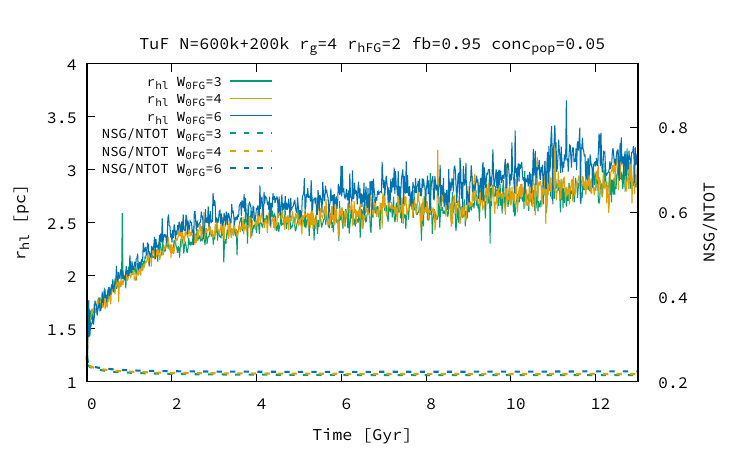}
  \hspace*{9cm}\includegraphics[width=0.49\linewidth,]{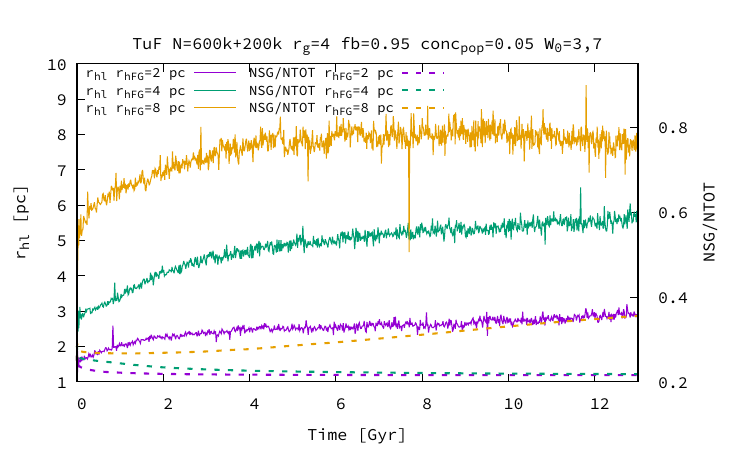}
  \caption{(contd.) Every row, starting from the top, differ in: \CONCPOP -- concentration parameter, \Wofg -- King parameter for FG, and \rhfg [pc] -- half-mass radii for FG (thus, only TuF models).
  }
  \label{f:2:NSG2}
\end{center}
\end{figure*}


The second row in Figure~\ref{f:2:NSG1} shows \MOCCA models for different Galactocentric distances (\rg). TF models with larger \rg have the same initial masses, but larger \rtid. All models rapidly lose FG stars during the early evolution dominated by the cluster's expansion which is driven by mass loss associated with stellar evolution; this results in a significant increase of \NSGNTOT approximately independent of the Galactocentric distance. This result is consistent with the findings of \citet{Vesperini2021MNRAS.502.4290V} who found that the final \NSGNTOT is mainly determined by the early evolutionary phases while the effects of Galactocentric distances playing only the role of a much less important second parameter. 

For our model with \rg = 2~kpc, the final \NSGNTOT is smaller than for other models only because this cluster completely dissolves at the end of the simulation and the final stages of evolution are dominated by the mass loss from the center by dynamical interactions, preferably involving SG stars. 

For TuF models the \NSGNTOT practically does not significantly depend on \rg (from \rg=~2~kpc to 8~kpc). As expected, strongly underfilling models do not lose a significant fraction of FG stars and the values of \NSGNTOT at the end of the simulation is similar to the initial one (they drop only slightly).

In the third row in Figure~\ref{f:2:NSG1} there are models with different binary fractions (fb). Models with different fb have the same total mass. TF models with higher fb have only slightly larger final \NSGNTOT. This is connected with faster mass segregation and larger energy generation in dynamical interactions by models with larger fb. Interestingly, it seems that fb = 0.1 (usually chosen in N-body simulations) result in a similar values for \NSGNTOT as for larger fb. For TuF models fb does not have any visible influence on \NSGNTOT and only slightly on \rhob. This is connected with the fact that clusters behave like an isolated models and the structure of the cluster controls the efficiency of the central energy source.

The fourth row in Figure~\ref{f:2:NSG1} illustrates the results obtained by assuming different values for the maximum stellar mass, \Mmax, for the SG. The general expectation for a single-population clusters is that a cluster with a larger \Mmax (or, more in general, with a larger initial fraction of massive stars) should undergo a stronger initial expansion because of larger amount of mass lost due to stellar evolution (see e.g. \citealt{Fukushige1995}). However, the effect of varying \Mmax for the SG in a multiple-population cluster is more complex. Detailed investigation of the evolution of models with different \Mmax values revealed very interesting dynamical effects associated with the change  in \Mmax. For the TF model with larger SG \Mmax, as expected, the first phase of the cluster evolution connected with the stellar evolution, leads to a larger mass loss and a more rapid increase of \NSGNTOT (effect more apparent for the case \rg~=~4~pc and \CONCPOP~=0~.1, not shown in this paper). During the subsequent long-term evolution, however, the increase of \NSGNTOT slows down and even slightly decreases. The model with smaller \Mmax, on the other hand, continues its evolution towards larger values of \NSGNTOT.

The different dynamical evolution of the two systems with different SG \Mmax can be explained by the underlying differences in the population of stellar mass BHs in the two systems. The model with SG \Mmax~=~20 \msun does not create a population of SG BHs which are instead produced in the system with SG \Mmax~=~150 \msun. Thus, from the very early phases of its evolution the model with the larger SG \Mmax creates dense core with a population of stellar mass BHs which act as an energy source. The model with SG \Mmax~=~20 \msun, on the other hand, does not have very massive stars (and the BHs they form) in the centrally concentrated SG subsystem. BHs and massive stars have to segregate in the central regions from the more extended and less concentrated FG system. This process needs some time. It needs $\sim$ 2 Gyr to mass segregate 70\% of BH from FG. For the model with \Mmax~=~150 \msun it is needed only 1 Gyr.

After mass segregation FG BHs will start to generate energy and support the cluster evolution. The longer timescale of mass segregation for the system with the smaller SG \Mmax leads to a stronger early expansion and to different spatial structures (Lagrangian radii for \Mmax~=~20 \msun are larger than for \Mmax~=~150 \msun). The more extended structure of the systems with \Mmax~=~20 \msun leads to stronger mass loss of FG stars and larger \NSGNTOT and, more in general, to a more rapid cluster dissolution. The same process is much more profound and better visible for the case \rg~=~4~pc and \CONCPOP= 0.1 (not shown in the paper), for which the model with \Mmax~=~20 \msun presents even larger increase of \rhob, and dissolves 4 Gyr faster than the model with \Mmax~=~150 \msun.

A similar behavior and differences between models with different values of the SG \Mmax, although to a much smaller extent, are found for the TuF models. In the case of TuF models both models survive until the Hubble time too.

The fifth row in Figure~\ref{f:2:NSG2} presents TF, and TuF models for different values of the concentration parameter (\CONCPOP) that determines the relative spatial distributions of the FG and SG populations (see Section \ref{s:NumSimul}). For the TF models \NSGNTOT rapidly increases during the cluster early evolution for all the values of \CONCPOP; the values of \NSGNTOT at the end of the early evolution (at t $\sim$ 1-2 Gyr) is slightly larger for lower values of \CONCPOP for which the larger concentration of the SG population leads to a stronger preferential loss of FG stars. The larger central densities of models with smaller values of \CONCPOP affect also the long-term evolution and the subsequent evolution of \NSGNTOT but the general trend between \NSGNTOT and \CONCPOP is set already during the cluster's early evolutionary phases. 

For TuF models the loss of FG stars and the ensuing variation of \NSGNTOT are much milder and again the final values for \NSGNTOT are just slightly lower than the initial ones. Also for the models presented in these panels the final values of \rhob fall within the range of those found for Galactic globular clusters.

The sixth row in Figure~\ref{f:2:NSG2} shows the models for different \Wofg which has the most profound influence on the \NSGNTOT ratio and dissolution time (\tdiss). For TF models, the ratio of the FG half-mass to tidal radius increases for decreasing values of the \Wofg values; this trend implies that GCs with smaller \Wofg values lose FG stars more efficiently during the early cluster's expansion leading to a more significant increase of \NSGNTOT.

The process becomes less efficient for larger \Wofg and for \Wofg = 6 \NSGNTOT increases only slightly in comparison to the initial values. This confirms earlier findings by \citet{Vesperini2021MNRAS.502.4290V} that cluster models for which \Wofg is too large (larger than 6) will not produce clusters with large present-day \NSGNTOT. We point out, however, that the inclusion of additional dynamical processes such as primordial gas expulsion and early tidal shocks might affect the structure of the clusters and lead to the efficient loss of FG stars also for larger initial values of \Wofg.  For TuF models, the dependence of cluster parameters and \NSGNTOT on \Wofg is practically negligible. Independently of \Wofg clusters have a lot of space to expand up to \rtid. So, the mass loss is very similar.

The last row in Figure~\ref{f:2:NSG2} shows models for different \rhfg (only TuF models). The \rtid radius is the same for all the models. These models nicely show that for larger \rhfg values and constant \rtid the ratio \rtid/\rhfg is increasing and coming closer to TF. Therefore more and more FG stars escape which leads to increase of the ratio. Since all these models are tidally underfilling, GCs can undergo their initial expansion without immediately losing FG stars and thus \NSGNTOT does not increase as much as TF models. Also in this case, the inclusion of the dynamical processes mentioned above (gas expulsion, early tidal shocks) may lead to a more significant loss of FG stars and increase of the \NSGNTOT ratio.

For TuF models, we find an early slight decrease (about <~1~\%) in \NSGNTOT ratio during the initial $\sim$~100 Myr. This is the result of the preferential ejection of SG stars and binaries due to dynamical interactions in the innermost regions (the main star loss mechanism in TuF clusters) where the SG is the dominant population.

\subsection{The role of initial dynamical parameters on the evolution of clusters and their multiple stellar populations}

As shown in the numerous past theoretical studies on the evolution of globular clusters, the  dynamical history of these systems depends on various internal dynamical parameters as well as on the properties of the external tidal field of their host galaxies (see e.g. \citealt{Heggie2003gmbp.book.....H}). The presence of MSP with different dynamical properties significantly broadens the parameter space describing the possible initial properties of globular clusters and their subsequent evolution. The small survey of simulations carried out for this investigation allows us to start building a comprehensive picture of the possible dynamical paths followed by multiple-population clusters; although a much larger number of simulations will be necessary to build a more complete picture, the simulations presented here provide a number of key indications on the role played by some of the parameters describing their initial structural properties.

Table~\ref{t:Trends} summarizes the results emerging from Figure~\ref{f:2:NSG1} and illustrate the role played by various parameters  in determining the variation of \NSGNTOT, cluster mass at the Hubble time, or dissolution time. As expected \tdiss increases as  N (for TF, and TuF), \rg, \Wo and \Mmax for TF models increase. In turn, in order to increase \NSGNTOT ratios one can increase fb for TF models, and \rhfg for TuF models.

\begin{table*}
\caption[Trends of \NSGNTOT ratios]{Top table summarizes how by increasing one parameter (1st column) one can influence \NSGNTOT ratios, M -- total GC mass, and \tdiss -- time of dissolution of GC for both tidally filling (TF, 2nd column) and tidally underfilling models (TuF, 3rd column). The arrows show whether a given parameter is increasing (\up), decreasing (\down), particularly strong (\upstrong, \downstrong), or weak (\upweak, \downweak). The bottom table shows the same but from the point of view of the GC global parameters -- which initial parameters one has to increase to get higher \NSGNTOT, M, or \tdiss values (up to Hubble time).}
\centering
\begin{tabular}{|c| c |c|}
\hline 
Parameter & TF & TuF\\ 
\hline\hline
 $\rm N$ & \NSGNTOT \same, M \upstrong, \tdiss \upstrong & \NSGNTOT \same, M \up, \tdiss \upstrong \\ 
 \rg & \NSGNTOT \same, M \upstrong, \tdiss \up & \NSGNTOT \same, M \upstrong, \tdiss \same \\
 fb & \NSGNTOT \upweak, M \upweak, \tdiss \downstrong & \NSGNTOT \same, M \same, \tdiss \same \\
 \Mmax  & \NSGNTOT \down, M \same, \tdiss \up & \NSGNTOT \same, M \same, \tdiss \same \\
 \CONCPOP  & \NSGNTOT \down, M \same, \tdiss \same & \NSGNTOT \same, M \upweak, \tdiss \same \\
 \Wofg & \NSGNTOT \downstrong, M \upstrong, \tdiss \upstrong & \NSGNTOT \same, M \same, \tdiss \same \\
 \rhfg & & \NSGNTOT \up, M \down, \tdiss \same \\
 \hline
\end{tabular}

\begin{tabular}{|c| c |c|}
\hline 
Parameter & TF & TuF\\ 
\hline\hline
 \tdiss \up & N \upstrong, \rg \up, \Wofg \upstrong, \Mmax \up & N \upstrong \\ 
 \NSGNTOT \up & fb \upweak & \rhfg \up\\
 M \up & N \upstrong, \rg \upstrong, fb \up, \Wofg \upstrong & N \up, \rg \upstrong\\
 \hline
\end{tabular}
\label{t:Trends}
\end{table*}

\section{Discussion}
\label{s:Discussion}

The main goal of this study is to provide an initial exploration of the role of various parameters in determining the evolution of a few key dynamical properties of globular clusters and those of their multiple stellar populations. Such exploration is the first step towards the identification of the regions of the initial parameters space leading to properties generally consistent with the present-day observed properties of Galactic globular clusters.
In the following sections we further discuss some of the results of the simulations  introduced in this paper. In future studies we will extend our survey of simulations and carry out an investigation aimed at modeling the evolution of populations of globular clusters.

\subsection{Distribution of SG stars in the cluster}\label{subsec:conc-radius-vs-ratio}

We start our discussion by focusing our attention on the radial variation of the fraction of SG stars. As discussed in the Section~\ref{s:Intro}, several formation models and hydrodynamical simulations of SG formation predict that SG stars form more centrally concentrated in the inner regions and, during the subsequent early and long-term dynamical evolution, the strength of the initial spatial differences gradually decreases. Some clusters may retain some memory of these initial differences while in others complete mixing may be reached. Observational studies often estimate \NSGNTOT in a limited portion of the cluster and it is therefore important to establish how these estimates may be affected by radial variations of \NSGNTOT.

\begin{figure}
\begin{center}
  \includegraphics[width=0.99\linewidth]{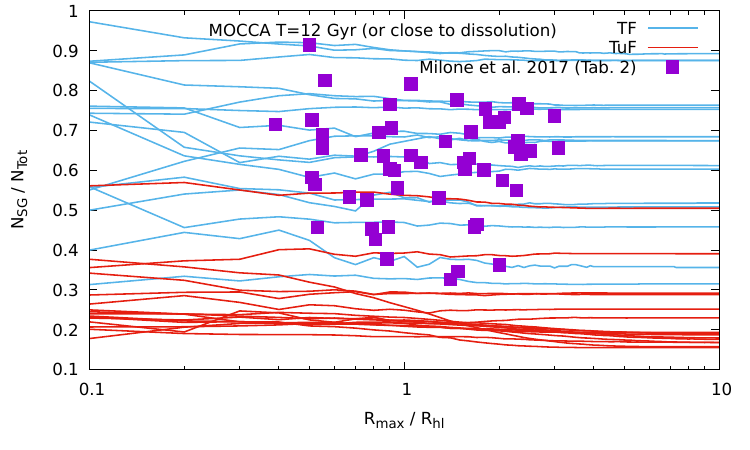}
  \caption{Milky Way GCs coverage with \MOCCA models for 2D \NSGNTOT \citep{Milone2017MNRAS.469..800M} ratios computed as a cumulative value for different radii (scaled to \rhob). Every line correspond to one \MOCCA simulation. The selected \MOCCA models are the same as for Figure~\ref{f:2:NSG1}. \NSGNTOT ratios are colored by initially being TF or TuF. Only for couple of models there is a large difference for \NSGNTOT ratios between small and large radial distances (for the one which are closer to dissolution). For majority of the models there is no significant different for \NSGNTOT ratios between measuring it in the very center or around \rhob radius. For details see Section~\ref{s:Discussion}.
  }
  \label{f:5:NSGall}
\end{center}
\end{figure}

Figure~\ref{f:5:NSGall} presents the \NSGNTOT ratio of MW GCs and the cumulative \NSGNTOT ratios computed for different radii in the \MOCCA models. Each line corresponds to one \MOCCA simulation at 12 Gyr (for models dissolving earlier which did not survive to this time, it is the profile calculated from the last saved snapshot which still holds at least 1\% of the initial mass). The \MOCCA models are those presented in Figure~\ref{f:2:NSG1}. The cumulative \NSGNTOT ratios are computed for every \MOCCA simulation for $\rm R_{max}$ from 0.1~\rhob, up to 10~\rhob. This figure clearly shows that despite the broad range of initial conditions explored and the fact that the models in our survey reach a variety of different degrees of spatial mixing, the values of \NSGNTOT measured within clustercentric distances similar to those usually covered by observational studies are representative of the global values. Moreover, as already discussed in the previous section, this figure further illustrates a clear dichotomy between initially TuF and TF models where the latter are generally characterized by larger final values of \NSGNTOT and that the values of \NSGNTOT found in out TF models falls  within the range of those found in observations of Galactic GCs.

\subsection{On the range of values of \NSGNTOT}
\label{subsec:conc-comparison}

As shown in the various panels of Figure~\ref{f:2:NSG1}, the values of \NSGNTOT in our models can reproduce the range of values observed in Galactic globular clusters (see \citealt{Milone2017MNRAS.464.3636M}) while producing systems with half-light radii  also generally consistent with those observed. Further investigation extending the survey of simulations to include a broader range of initial masses will allow us to study a populations of clusters and explore the trends between \NSGNTOT and other cluster properties.

We generally excluded models that formed an IMBH from the analysis presented in this paper. However, some models do form them. We plan to study the formation and evolution of those IMBHs and their influence on the \NSGNTOT in the future work. In the newest \MOCCA version we have e.g. IMBH seed BHs being formed as a result of a runaway merging scenario. 

Our analysis shows that in order to reach values of \NSGNTOT consistent with those observed the FG populations must be initially TF or slightly TuF and be characterized by initial values of \Wofg smaller than about 5-6.  More extreme initial central concentration (i.e. larger values of \Wo) for the FG would lead to a modest early loss of FG stars and smaller increase of \NSGNTOT. Such values for \Wofg are in agreement with those expected from the process of residual gas removal after FG is formed. Sudden gas removal leads to a much shallower central potential and a more extended FG characterized by small \Wo \citep{Leveque2022gas}.

As for the requirement on the initial degree of tidal filling and the possible link between the initial and present-day structural properties, it is important to point out that as shown in this study and previously in \citet{Vesperini2021MNRAS.502.4290V}, most of the evolution of \NSGNTOT occurs in the cluster early evolutionary phases with a much more modest increase of \NSGNTOT during the subsequent long-term evolutionary phase; this implies that the relevant dynamical requirements for the evolution of \NSGNTOT are to be considered in the context of the strength of the tidal field during the first phase of a cluster's evolution. A number of studies (see e.g. \citealt{2022MNRAS.515.1065M}; see also \citealt{badry2016} for the possible effect of feedback-driven fluctuations in the gravitational potential of a galaxy in the early radial migration towards weaker tidal fields)  have suggested that clusters experience a stronger tidal field (and stronger time variations and tidal shocks) in the first 1-2 Gyr of their evolution and later migrate to larger galactocentric distances and weaker tidal fields. The idea that short time-scale cluster migration from gas-rich formation environments (via mechanisms like e.g. frequent galaxy mergers) has been proposed as a mechanism for the long-term survival of GC progenitors \citep{Kruijssen2015,Forbes2018}.

As for the evolution of multiple-population clusters, in addition to consequences associated to the possible role of additional processes contributing to the early loss of FG stars (e.g. early tidal shocks), this migration implies that the required condition of tidally filling clusters would correspond to clusters with a more compact structure and smaller half-mass radii than those needed to be TF in weaker tidal fields. As far as the evolution of the fraction of SG stars and the cluster's mass is concerned, this could be a plausible pathway to support a modified scenario in which a sizable number of extended FG stars can be lost due to strong tidal stripping in the first 1-2 Gyr of cluster evolution  in the initial stronger tidal field while the long-term evolution of the cluster's mass driven by two-body relaxation would proceed at a slower rate for clusters migrating into a weaker tidal field (see \citealt{Onorato2023} for a first investigation exploring the implications of such a transition for the evolution of the multiple populations in the massive Galactic cluster NGC~2419). The resulting masses, sizes and \NSGNTOT in such a scenario incorporating a transition from a stronger to a weaker tidal field may naturally lead to properties consistent with those observed in present-day clusters. Further investigation of this scenario is currently in progress and will be presented elsewhere.

\section{Conclusions and future work}
\label{s:Conclusions}

In this paper we have explored the evolution of multiple-population clusters for a broad range of initial conditions expanding those considered in previous studies. The exploration presented in this paper provided a more comprehensive picture of the evolution of a number of key properties of multiple-population clusters and will serve as the basis for future investigations and surveys aimed at building specific models for the properties of Galactic globular clusters and their multiple populations.

In our models we have considered different initial number of stars, initial concentration of the FG and SG populations, galactocentric distances, binary fractions, and upper mass limits on the initial mass function, as well as configurations in which the FG was tidally filling or tidally underfilling. This exploration has allowed us to start shedding light on how a number of parameters depends on the initial conditions adopted. In particular, in this paper we have started to explore how the clusters' lifetime, total mass and fraction of SG stars depend on the initial values assumed for those parameters and properties. The results are summarized in Table 2. In addition to the trends reported in Table 2, our conclusions can be summarized as follows.

\begin{itemize}
\item In agreement with previous studies, we find that in models starting with the FG tidally filling, \NSGNTOT can undergo a significant evolution reaching higher values falling in the range of those observed in Galactic globular clusters. Models with a FG initially tidally underfilling, on the other hand, do not lose a significant number of stars and retain values of \NSGNTOT similar to the initial ones. 
\item In order for the clusters to undergo a significant increase in the \NSGNTOT ratio, the initial spatial distribution of the FG population modeled as that of a King model must have an initial value of the central dimensionless potential $W_0 \sim 5-6$ or smaller. 
\item The \NSGNTOT ratio is changing most noticeably during the first 1-2 Gyr of the cluster's evolution and it does not change significantly during the subsequent evolution. The initial conditions and the environment in which a GC was born are thus likely to play crucial role in shaping the final values of \NSGNTOT ratios (see Figure 2).
\item In most of the models we have investigated in this paper, we find only mild differences ($<0.1$) between the value of \NSGNTOT calculated  within the inner regions (e.g. within 0.1 \rhob) and the values calculated within 1-2 \rhob. In most cases values of \NSGNTOT calculated within 1-2 \rhob (a radial range typical of many observational studies) are representative of the global values for the entire cluster (see Figure 3).
\item Many of the models and initial conditions explored in this paper produce final values of \NSGNTOT, masses, core and half-light radii overlapping with those observed  in Galactic globular clusters (see Figures 1 and 2) and our survey has shed light on the range of initial conditions resulting in properties generally consistent with observations. We point out, however, that the goal of this paper was not to produce a complete model for the Galactic globular cluster system and the trends observed for the properties of the multiple populations. Additional simulations including clusters even more massive than those considered here are necessary for a more comprehensive investigation.  Our future models will also include additional ingredients and refinements; in particular we will include the possible effects of  a tidal field  varying in time due to fluctuations in the cluster's birth environment as well as a result  the cluster's migration from the site of formation to various galactocentric distances, the dynamical effects associated to a delay between the time of FG and SG formation, and tidal effects due to eccentric orbits. A significant extension of the survey presented here including these effects and exploring a broader range of initial conditions will be presented in future papers.
\end{itemize}

\begin{acknowledgements}

This research has been partially financed by the Polish National Science Centre (NCN) grant 2021/41/B/ST9/01191. EV acknowledges support from NSF grant AST-2009193. AA acknowledges support for this paper from project No. 2021/43/P/ST9/03167 co-funded by the
Polish National Science Center (NCN) and the European Union Framework Programme for Research
and Innovation Horizon 2020 under the Marie Skłodowska-Curie grant agreement No.
945339. For the purpose of Open Access, the authors have applied for a CC-BY public copyright license to any Author Accepted Manuscript (AAM) version arising from this
submission. MOH acknowledges support by the Polish National Science Center grant 2019/32/C/ST9/00577. 

We thank the referee for all the comments and suggestions that helped us to improve the paper.

\end{acknowledgements}

\section*{Software}

\MOCCA code is open source\footnote{\url{https://moccacode.net/license/}} for our collaborators. We are open to start new projects, in which one could use already existing \MOCCA simulations, or start new ones.

\textsc{beans}\footnote{\url{https://beanscode.net/}} software is open source and it is freely available for anyone.





\section*{Data availability}
The data from this article can be shared on request.

%
%

\bibliographystyle{aa}
\bibliography{biblio}

\begin{appendices}
\newpage

\section{BEANS script}
\label{s:AppendixB}

We present one of the scripts which were used while working on this paper. The whole data analysis was done with \BEANS software (see Section~\ref{s:DataAnalysis}). We take this opportunity to show how one can analyze huge data sets (in our case astronomical data coming from numerical simulations) in easy way using Apache Pig scripts. Other scientists might be interested in using \BEANS for their research too.

For detail description about specific Apache Pig keywords and instructions one can find in \citet[Appendix B]{Hypki2022}, or in Apache Pig documentation\footnote{\url{https://pig.apache.org/docs/r0.17.0/index.html}}. It is adviced to read it first. Here, only the main steps of the scripts will be briefly described. The script computes cumulative profiles of FG and SG stars for all \MOCCA simulations (used in this paper) for a number of selected timesteps for which there are available snapshot data. The comments in Apache Pig scripts are the lines starting with '-~-' characters and they are used to describe code snippet below.

\begin{verbatim}
-- <1> Reading data from snapshot files from MOCCA
--     simulations, from the Survey5, for a few 
--     selected timesteps (0, 0.4, 1, 2, 5... Gyr)
snap = load 'datasets="mocca survey5" 
               tables="snapshot" 
               filter="(timenr == 0) 
               OR (tphys>400.0 tphys<430.0) 
               OR (tphys>1000.0 tphys<1030.0) 
               OR (tphys>2000.0 tphys<2030.0) 
               OR (tphys>5000.0 tphys<5030.0) 
               OR (tphys>7000.0 tphys<7030.0) 
               OR (tphys>10000.0 tphys<10030.0) 
               OR (tphys>12000.0 tphys<12030.0)"' 
               using BeansTable();

-- <2> Getting from snapshots only some selected 
--     columns: ID of the dataset (dsid), physical
--     time (tphys), radial distance (r), 
--     population ID (pop), and we create bins 
--     every r = 0.1 pc
snap = foreach snap generate 
            DSID(tbid)  as dsid,
            tphys,
            r, 
            popId1 as pop,
            FLOOR(r, 0.1) as bin;

-- <3> Compute total number of FG and SG stars
--     for every MOCCA simulation, for every 
--     timestep. First we group all data based
--     on these criteria (group ... by), and 
--     later we compute the numbers (foreach ...
--     COUNT(...)) in every group. It will be 
--     used to normalize the profiles later.
snapGr = group snap by (dsid, tphys, pop);
snapTotal = foreach snapGr generate
            group.$0            as dsid,
            group.$1            as tphys,
            group.$2            as pop,
            COUNT(snap.pop)     as popCount;

-- <4> Instructions in blocks <4>, <5> and <6>
--     concerns actual computation of the 
--     profiles. They were divided into blocks
--     for clarity. Block <4> groups all data
--     by bins (with width r = 0.1 pc). Under
--     the variable (alias) 'snapGr' all the data
--     are grouped by MOCCA simulations, physical 
--     times, population id, and later by bins.
--     This is needed for the next instruction,
--     where one can compute how many FG and SG 
--     are in every bin. 
snapGr = group snap by (dsid, tphys, pop, bin);
snap = foreach snapGr generate
            group.$0 as dsid,
            group.$1 as tphys,
            group.$2 as pop,
            group.$3 as r,
            COUNT(snap.r) as count;
            
-- <5> Creating self-join between the results of
--     the previous code (alias 'snap'). In the
--     first step it is created a copy of alias
--     alias 'snap'. In the next step the actual
--     self-join operator is used. Then, in this
--     joined alias the rows are filtered and
--     the one comming from the 'snap2' which have
--     larger radial distances are removed - this 
--     is needed in the last <6> step to compute 
--     cumulative values. 
snap2 = foreach snap generate *;
snapJoined = join snap by (dsid, tphys, pop), 
                snap2 by (dsid, tphys, pop);
snapJoined = filter snapJoined 
                by snap::r <= snap2::r;
snapJoined = foreach snapJoined generate 
                snap::dsid      as dsid,
                snap::tphys     as tphys,
                snap::count     as count,
                snap2::r        as r2,
                snap2::pop      as pop;

-- <6> Computing cumulative profiles based on the 
--     rows from the previous alias. The alias 
--     is first grouped by the radial distance 
--     'r2'. Then, one can compute how many rows 
--     there are inside such radius - these are 
--     in fact cumulative values for both 
-      populations FG, and SG.
snapJoinedGr = group snapJoined by (dsid, tphys, 
                pop, r2);
snap3 = foreach snapJoinedGr generate
                group.$0 as dsid,
                group.$1 as tphys,
                group.$2 as pop,
                group.$3 as r2,
                SUM(snapJoined.count) as r2Count;

-- <7> These two rows are simply joining the 
--     results of the previous rows (with 
--     cumulative values) with the rows which 
--     store total number of FG and SG stars 
--     (needed for normalization). 
snap3Joined = join snap3 by (dsid, tphys, pop), 
                snapTotal by (dsid, tphys, pop);
snap3Joined = foreach snap3Joined generate
                snap3::dsid         as dsid,
                snap3::tphys        as tphys,
                snap3::pop          as pop,
                snap3::r2           as r,
                snap3::r2Count      as count,
                snapTotal::popCount as popCount;

-- <8> Storing the results in BEANS. 
store snap3Joined into 
                'name="snap cumulative survey5"' 
                using BeansTable();
\end{verbatim}



\end{appendices}

\end{document}